\documentclass[aps,prb,twocolumn,floatfix,10pt]{revtex4}

\newcommand{\bk}{{\bf k}}

\newcommand{\be}{\begin{equation}}
\newcommand{\ee}{\end{equation}}
\newcommand{\bea}{\begin{eqnarray}}
\newcommand{\eea}{\end{eqnarray}}

\newcommand{\bwt}{\begin{widetext}}
\newcommand{\ewt}{\end{widetext}}

\usepackage{float}
\usepackage{graphics,graphicx,amsmath}
\usepackage{tabularx,float}
\usepackage{epsfig}

\usepackage{bm,bbold}
\usepackage{color}
\usepackage{verbatim}

\usepackage[T1]{fontenc}
\usepackage{libertine}

\begin{document}

\title{Electronic properties of single-layer and multilayer transition metal dichalcogenides {\em MX}$_2$ ($M=$ Mo, W and $X=$ S, Se)}

\author{R. Rold\'an,$^1$ J.A. Silva-Guill\'en,$^2$ M.P. L\'opez-Sancho,$^1$ F. Guinea,$^1$ E. Cappelluti$^3$ and P. Ordej\'on$^2$}
\address{$^1$Instituto de Ciencia de Materiales de Madrid,
CSIC, Sor Juana Ines de la Cruz 3, 28049 Cantoblanco, Madrid, Spain}
\address{$^2$ICN2 - Institut Catala de Nanociencia i Nanotecnologia, Campus UAB, 08193 Bellaterra, Spain}
\address{CSIC - Consejo Superior de Investigaciones Ciaentificas, ICN2 Building, 08193 Bellaterra, Spain}
\address{$^3$Istituto de Sistemi Complessi, U.O.S. Sapienza, CNR, v. dei
Taurini 19, 00185 Roma, Italy}

\begin{abstract}

Single- and few-layer transition metal dichalcogenides have recently emerged as a new family of layered crystals with great interest , not only from the fundamental point of view, but also because of their potential application in ultrathin devices. Here we review the electronic properties of semiconducting $MX_2$, where $M=$Mo or W and $X=$ S or Se. Based on of density functional theory calculations, which include the effect of spin-orbit interaction, we discuss the band structure of single-layer, bilayer and bulk compounds. The band structure of these compounds is highly sensitive to elastic deformations, and we review how strain engineering can be used to manipulate and tune the electronic and optical properties of those materials. We further discuss the effect of disorder and imperfections in the lattice structure and their effect on the optical and transport properties of $MX_2$. The superconducting transition in these compounds, which has been observed experimentally, is analyzed, as well as the different mechanisms proposed so far to explain the pairing. Finally, we include a discussion on the excitonic effects which are present in these  systems. 

\end{abstract}

\date{\today}
\maketitle

\section{Introduction}

Two-dimensional layered materials are currently being object of great attention
due to their  physical properties. Single-layer and few-layer graphene have
received much attention and bolstered this field of research.\cite{CastroNeto_etal} 
Recently, the focus is widening to other two-dimensional materials with
interesting  structural and electronic properties.\cite{BP12}
Transition metal dichalcogenides (TMD) form a new family of layered materials
that can be easily exfoliated and present promising electrical and optical properties.\cite{NG05}
Among these materials, semiconducting TMD are of special interest since
the gap  present in both single-layer and multi-layer samples
makes them candidates for device applications.\cite{WS12,JH14,GZ14}

The band structure of those compounds dramatically changes from single-layer to multi-layer samples, involving a transition from a
direct gap for single-layer samples to an indirect gap for multi-layer
samples,\cite{CG13} as it has been observed experimentally,\cite{MH10,ZE13,JO13,ZS14} 
pointing out the important role of interlayer coupling.\cite{CR13} Furthermore, their electronic properties are highly sensitive to the external conditions such as temperature, pressure or strain. For instance, an insulator/metal transition
can be induced under particular conditions.\cite{FL12,LZ12,PZ12,PV12,SS12,SS12b,YL12,LA13,GH13,SY13,HT13,HP13,CS13}
This tunability of the  gap is very interesting for optoelectronic
applications.\cite{WS12}

Other important feature of the TMD is the possibility to control quantum degrees of freedom as the electron spin, the valley pseudospin and layer pseudospin.\cite{XH14} In fact, the strong spin-orbit interaction in these compounds, and the coupling of the spin, the valley and the layer degrees of freedom opens the possibility to manipulate them  for future applications in spintronics and valleytronics devices.\cite{MH12,MS13,CF12,SU12,XY12,WX13,ZC12,OR13} 
The  spin orbit coupling (SOC)
lifts the spin degeneracy of the energy bands in single layer samples
due to the abscence of inversion symmetry.\cite{ZCS11}
By time reversal symmetry the spin splitting in inequivalent valleys must be opposite,
leading to the so called spin-valley coupling,\cite{XY12}
which have been observed experimentally\cite{CF12,ZC12,MH12,WX13,WS13,ZC13}
and have also been studied theoretically.\cite{FX12,SX13,RMA13,RGP13,LX13,RG14}

 Due to the reduced dielectric screening in monolayer and few-layer samples of TMDs, excitonics effects are especially relevant in these compounds. The existence of neutral and charged excitons, as well as their possible manipulation for optoelectronic applications, is attracting a lot of interest from both, experimental\cite{Ross_2013,MS13,WL13,Jones_2013} and theoretical\cite{R12,CL12,BHR13} points of view.

Similarly as in graphene, metallic behavior can be induced in semiconducting TMD by means of electric field effects or by doping. At high carrier concentrations ($n \sim 10^{14}$ cm$^{-2}$), and in the presence of high-$\kappa$ dielectrics, MoS$_2$ becomes superconductor, with a doping-dependent critical temperature T$_c(n)$ which exhibits a {\it superconducting  dome} with a maximum for a certain range of doping $n$ and drops to zero at sufficiently large values of $n$,\cite{TT12,YI12} and whose understanding is focus of a number of recent theoretical works.\cite{RCG13,RHW14,YML14,MRA14}

Another topic of current research is the effect of disorder on the optical and electronic properties of TMDs. In particular, the presence of vacancies or adatoms in the samples can modify  their mobility, and the importance of short-range disorder is thought as one of the main limitations for the mobility of chemical vapor deposition (CVD) grown single-layer MoS$_2$ \cite{ZA14,SE14}.
Theoretically, this problem has been studied using {\it ab initio} methods \cite{AC11,MH11,WP12,KK12,GH13b,ZZ13,ES13,LR13}, as well as  {\it real space} tight-binding (TB) methods which can simulate realistic samples.\cite{YG14}

The present paper analyzes the electronic properties of the
group-VIB {\em MX}$_2$ (where $M=$ Mo, W and $X=$ S, Se). The corresponding band structure for single layer, bilayer and bulk compounds is obtained from density function theory (DFT) calculations, which includes spin-orbit interaction. 
We discuss the role played by both, the metal and the chalcogen atoms, in the optical and electronic properties of these materials. We review the effects of strain and disorder in the electronic spectrum, as well as the  superconducting  transition in highly doped samples.

\section{Electronic structure}

\begin{figure}[t]
\includegraphics[scale=0.37,clip=]{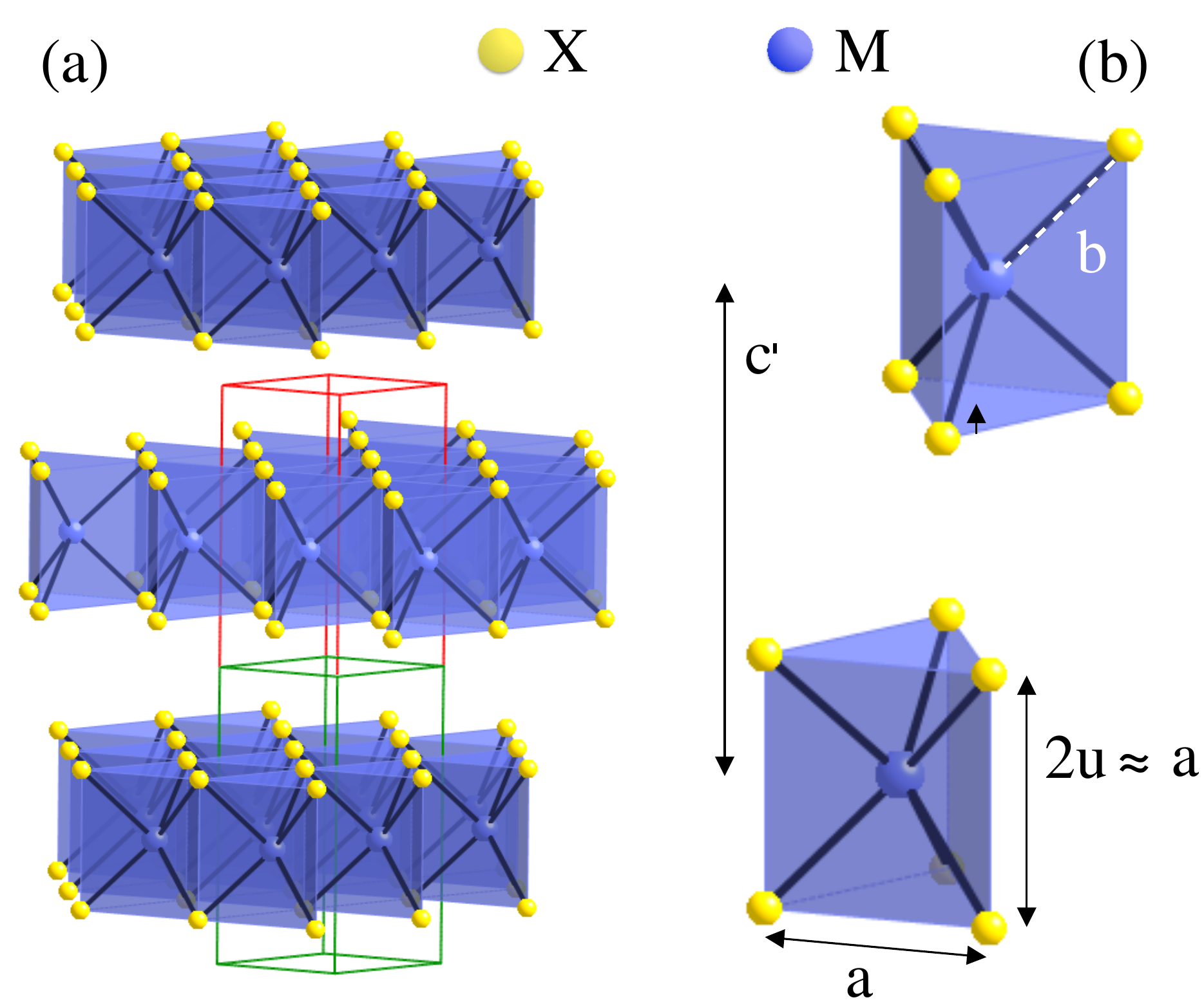}
\caption{(a) Schematic representation  of the atomic structure of {\em MX}$_2$.
The bulk compound has a 2H-{\em MX}$_2$ structure
with two {\em MX}$_2$  layers per unit cell, each layer
being built up from a trigonal prism coordination unit.
The small green rectangle represents the unit cell of a monolayer of {\em MX}$_2$,
which is doubled (red extension) in the bulk crystal.
(b) Detail of the trigonal prisms for the two layers in the bulk compound,
showing the lattice constants and the definition
of the structural angles used in the text. 
}
\label{Fig:Structure}
\end{figure}

In this section we discuss the main features of the electronic band structure of single-layer, bilayer and bulk TMDs. For this, we will use  DFT calculations,
including the intrinsic spin-orbit interaction term for all atoms. The crystal structure of single-layer and multilayer  {\em MX}$_2$
is schematically shown in Fig.  \ref{Fig:Structure}. The basic unit block of {\em MX}$_2$ is composed of an inner layer of $M$ atoms
on a triangular lattice sandwiched between two layers
of $X$ atoms lying on the triangular net of alternating
hollow sites. We denote\cite{CG13} $a$ as the distance between
nearest neighbor in-plane $M-M$ and $X-X$ distances,
$b$ as the nearest neighbor $M-X$ distance and
$u$ as the distance between the $M$ and $X$ planes.
The {\em MX}$_2$ crystal forms an almost perfect trigonal prism
structure with $b$ and $u$ very close to the their ideal values
$b \simeq \sqrt{7/12}a$ and $u \simeq a/2$.
The experimental values of these lattice distances of the bulk
compounds are given in Table \ref{Tab:LattParam} for
the four materials investigated.\cite{LX13,KGF13,KA12}
The in-plane Brillouin zone is thus characterized by the
high-symmetry points $\Gamma=(0,0)$, K$=4\pi/3a(1,0)$,
and M$=4\pi/3a (0,\sqrt{3}/2)$.

\begin{table}
\begin{tabular}{|l|c|c|c|}
\hline
\hline
 & $a$ & $u$ & $c'$ \\
\hline
MoS$_2$ & $3.160$ & $1.586$ & $6.140$ \\
\hline
WS$_2$ & $3.153$ & $1.571$ & $6.160$ \\
\hline
MoSe$_2$ & $3.288$ & $1.664$ & $6.451$ \\
\hline
WSe$_2$ & $3.260$ & $1.657$ & $6.422$ \\
\hline
\hline
\end{tabular}
\caption{
Lattice parameters used for DFT calculation for  $MX_2$, as taken from
Refs. \onlinecite{BMY72}, \onlinecite{schutte} and \onlinecite{KA12}. $a$ represents the $M$-$M$ atomic distance,
$u$ the internal vertical distance between the $M$ plane
and the $X$ plane, and $c'$ the distance between the $M$ layers.
In bulk systems the $z$-axis lattice parameter is given by $c=2c'$.
All values are in \AA\, units.
}
\label{Tab:LattParam}
\end{table}

\begin{table}
\begin{tabular}{lccccc}
\hline
\hline
 & Bandgap & VB(K) & VB($\Gamma$) &  CB(K)& CB(Q)  \\
\cline{2-6}
  &\multicolumn{5}{c}{Monolayer}\\
\cline{2-6}
MoS$_2$  & 1.715 & 0.153 & & 0.0041 & 0.0832 \\
WS$_2$   & 1.659 & 0.439 & & 0.0333 & 0.339  \\
MoSe$_2$ & 1.413 & 0.193 & & 0.0258 & 0.0    \\
WSe$_2$  & 1.444 & 0.439 & & 0.0396 & 0.275  \\
\hline
\hline
  &\multicolumn{5}{c}{Bilayer}\\
\cline{2-6}
MoS$_2$  & 1.710-1.198 & 0.181 & 0.737 & 0.0    & 0.457 \\
WS$_2$   & 1.658-1.338 & 0.451 & 0.677 & 0.0357 & 0.528 \\
MoSe$_2$ & 1.424-1.194 & 0.213 & 0.649 & 0.0253 & 0.417 \\
WSe$_2$  & 1.442-1.299 & 0.454 & 0.649 & 0.0428 & 0.522 \\
\hline
\hline
  &\multicolumn{5}{c}{Bulk}\\
\cline{2-6}
MoS$_2$  & 1.679-0.788 & 0.245 & 1.018 & 0.0    & 0.874 \\
WS$_2$   & 1.636-0.917 & 0.482 & 1.426 & 0.044  & 0.922 \\
MoSe$_2$ & 1.393-0.852 & 0.267 & 0.695 & 0.0228 & 0.819 \\
WSe$_2$  & 1.407-0.910 & 0.504 & 1.075 & 0.0548 & 0.919 \\
\hline
\hline
\end{tabular}
\caption{Band gap and splitting of the valence band (VB) and conduction band (CB) at different points of the BZ obtained from our DFT calculations Figs. \ref{Fig:1L}-\ref{Fig:Bulk}. The gap for monolayer samples is always direct, whereas for bilayer and bulk systems we give the direct/indirect gap for each compound (which values are shown are separated by a hyphen as follows: direct-indirect). All the energies are expressed in eV. }
\label{Tab:OurDFT}
\end{table}

\begin{figure}[t]
\includegraphics[scale=0.15,clip=]{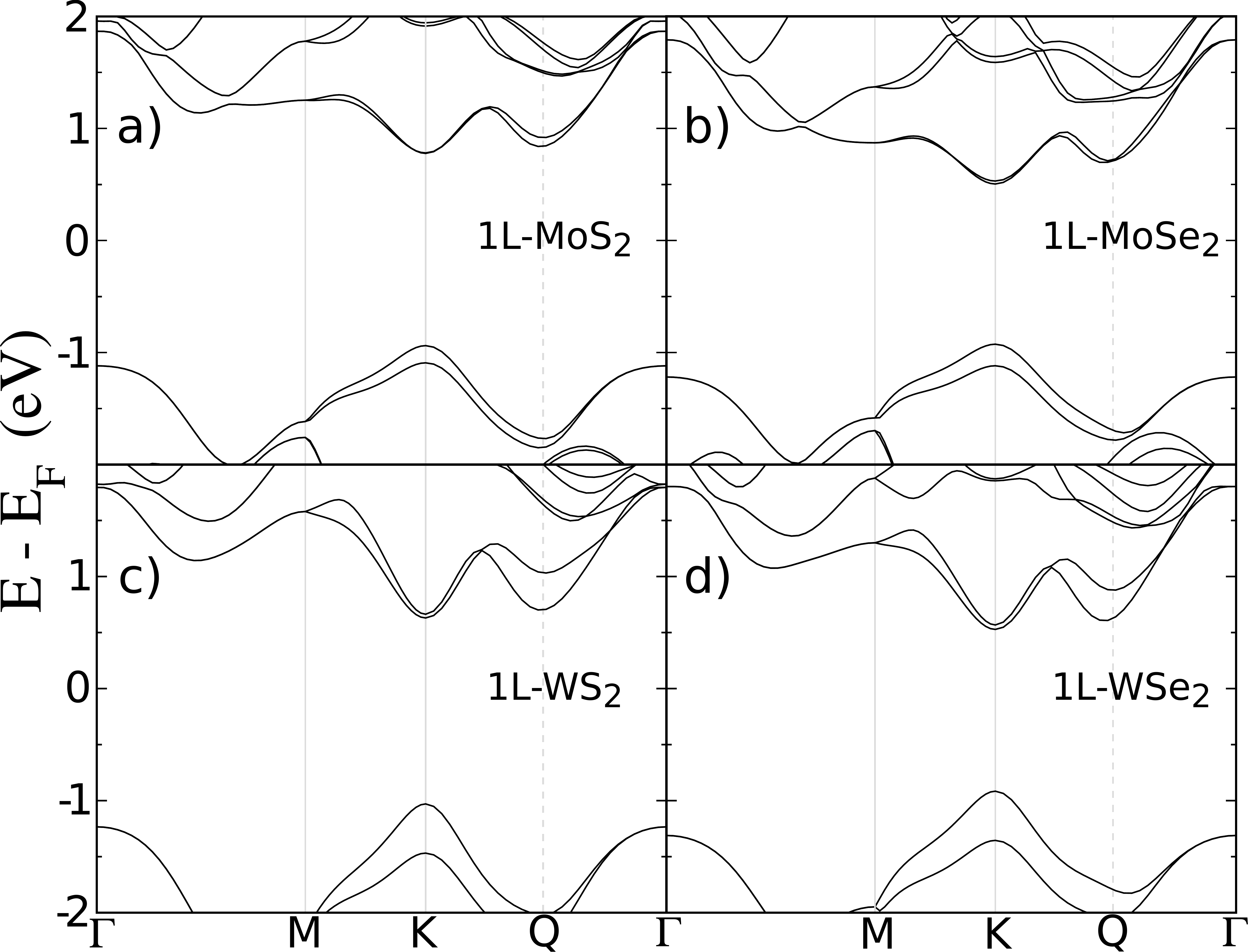}
\caption{Band structure of single layer MoS$_2$, MoSe$_2$, WS$_2$ and WSe$_2$
obtained from DFT calculations, including SOC. Dashed vertical lines indicate the position of the Q point in the BZ (see text).
}
\label{Fig:1L}
\end{figure}

DFT calculations are done using
the \textsc{Siesta} code,\cite{SS02,AS08} with the exchange-correlation potential of Ceperly-Alder\cite{CA80}
as parametrized by Perdew and Zunger.\cite{PZ81}
We consider also a split-valence double-$\zeta $ basis set including
polarization functions.\cite{AS99} 
The energy cutoff and the Brillouin zone sampling were chosen to converge the total energy with a value of 300 Ry and $30\times30\times1$ ($30\times30\times30$) in the case of the monolayer and bilayer (bulk), respectively. The energy cutoff was set to $10^{-6}$ eV.
We account for the spin-orbit interaction of the different compounds following the method developed in Ref. \onlinecite{FF06}
Figs. \ref{Fig:1L}-\ref{Fig:Bulk} show the obtained band structures for single-layer, bilayer and bulk samples of $MX_2$. 
The lattice parameters used in this calculation are given 
in Table \ref{Tab:LattParam}. In Table \ref{Tab:OurDFT}  we summarize the most important features of the band structures calculated in this work, like the energy gaps and the splitting of the valence and conduction bands at different points of the BZ. In Tables \ref{Tab:OthersDFT} and \ref{Tab:OthersDFTSplit} we list details of the band structure of different TMDs obtained in other works in the bibliography. A detailed description of the orbital character of the bands can be found in Ref. \onlinecite{CG13}. The valence and conduction bands are made by hybridization of the $d_{xy}$, $d_{x^2-y^2}$ and $d_{3z^2-r^2}$ orbitals of the transition metal $M$, and the $p_x$, $p_y$ and $p_z$ orbitals of the chalcogen atoms $X$. All the single-layer compounds shown in Fig. \ref{Fig:1L} are direct gap semiconductors, with the gap lying at the two inequivalent K points of the hexagonal BZ. The most important orbital contribution at the edge of the valence band at the K point is due to a combination of $d_{xy}$ and $d_{x^2-y^2}$ of the metal $M$, which hybridize to $p_x$ and $p_y$ orbitals of the chalcogen atoms $X$. On the other hand, the edge of the conduction band has a main contribution due to $d_{3z^2-r^2}$ of $M$, plus a minor contribution of $p_x$ and $p_y$ orbitals of $X$.\cite{CG13} 

\begin{table*}
\begin{tabular}{lccccc}
\hline
\hline
         & LDA    & GGA    & HSE    & GW     & Exp \\
\cline{2-6}
           & \multicolumn{5}{c}{Monolayer} \\
\hline           
MoS$_2$  
& 1.79\cite{KH12}/1.89\cite{KA12}/1.76\cite{HL14}/1.86\cite{DT11}
& 1.60\cite{R12}/1.59\cite{HL14}/1.67\cite{DT11} 
& 2.05\cite{R12}/2.32\cite{ELS11}/1.786\cite{ZG13}/2.25\cite{DT11}
& 2.82\cite{R12}/2.759\cite{CL12}/2.41\cite{MW13}/2.66\cite{DT11} 
& 1.8\cite{SP13,SW10}/1.90\cite{MH10}/1.86\cite{EM14}\\ 
WS$_2$   
& 2.05\cite{KA12}/1.94\cite{DT11}
& 1.56\cite{R12}/1.80\cite{MH11}/2.1\cite{KZH11}/1.9 
& 1.87\cite{R12}?/1.81\cite{DT11}
& 2.88\cite{R12}/2.32\cite{DT11}
& 2.91\cite{DT11} \\
MoSe$_2$ 
& 1.58\cite{KA12}/1.63\cite{DT11}
& 1.35\cite{R12}/1.44\cite{MH11}/1.44\cite{DT11}
& 1.75\cite{R12}/1.99\cite{DT11}
& 2.41\cite{R12}/2.31\cite{DT11}
& 1.76\cite{LE09}/1.58\cite{ZS14}/1.55\cite{TW12}\\
WSe$_2$  
& 1.61\cite{KA12}/1.45\cite{HL14}/1.74\cite{DT11}
&1.19\cite{R12}/1.25\cite{SP13}/1.32\cite{ZC13}/1.19\cite{HL14}
/1.55\cite{DT11}
& 1.68\cite{R12}/2.10\cite{DT11}
& 2.42(Indirect!)\cite{R12}/2.51\cite{DT11}
& 1.64\cite{SP13}/1.64\cite{ZC13}\\
\hline
\hline
           & \multicolumn{5}{c}{Bilayer} \\
\cline{2-6}
MoS$_2$  
& 1.68\cite{MW13}/1.75-1.17\cite{HL14}
& 1.275\cite{RNT11}/1.56-1.17\cite{HL14}
& 1.9\cite{ELS11}/1.480\cite{ZG13}($\Gamma$ to K)    
& 1.888\cite{CL12}/2.32\cite{MW13}/1.18($\Gamma$ to K)\cite{ZE13} 
& 1.88/1.6\cite{MH10}/1.51\cite{ZE13}  \\
WS$_2$   
& - 
& 1.37\cite{RNT11} 
& - 
& - 
& 1.7\cite{ZE13}  \\
MoSe$_2$  
&    
& 1.12\cite{RNT11}  
&   
&   
&   \\
WSe$_2$   
&1.35\cite{KA12}/1.68-1.27\cite{HL14}  
&1.23\cite{SP13}/1.1\cite{ZE13}/1.41-1.12\cite{ZC13}/1.19-1.14\cite{HL14}  
& -   
& -
& 1.5\cite{ZE13}/1.59-1.51\cite{ZC13}\\
\hline
\hline
           & \multicolumn{5}{c}{Bulk} \\
\cline{2-6}
MoS$_2$  
& 1.67\cite{MW13}/0.75\cite{KA12}/1.71-0.72\cite{HL14}
& 1.55-0.92\cite{HL14}/1.62-0.88\cite{J12}  
& 1.37\cite{ELS11}/1.328\cite{ZG13}
& 1.287\cite{CL12}/2.23\cite{MW13}/2.07-1.23\cite{J12}    
& 1.8/1.29\cite{MH10}/1.23\cite{KP82}/1.29\cite{BP01}  \\
WS$_2$  
& 0.89\cite{KA12} 
& 1.21\cite{SP13}/1.65-0.94\cite{J12}    
& - 
& 2.13-1.30\cite{J12}   
& 1.57\cite{SP13}/1.35\cite{KP82} \\
MoSe$_2$ 
& 0.80\cite{KA12}  
& 1.39-0.84\cite{J12} 
& - 
& 1.83-1.11\cite{J12}  
& 1.09\cite{KP82}/1.1\cite{BP01} \\
WSe$_2$  
& 0.97\cite{KA12} 
& 1.5-0.81\cite{HL14}/1.33-0.92\cite{J12}  
& - 
& 1.75-1.19\cite{J12} 
& 1.2\cite{KP82}/1.60-1.44\cite{ZC13} \\
\hline
\end{tabular}
\caption{Band gap obtained in other works using different methods or functionals as local-density approximation (LDA), generalised gradient approximation (GGA), hybrid functionals as the Heyd-Scuseria-Ernzerhof (HSE) and the GW approximation. Values separated by a hyphen are the direct-indirect gap. All the energies are expressed in eV.}
\label{Tab:OthersDFT}
\end{table*}

The main difference between the Mo$X_2$ [Fig. \ref{Fig:1L}(a) and (b)] and W$X_2$ [Fig. \ref{Fig:1L}(c) and (d)] compounds is observed in the splitting of the valence band for each case, which is due to SOC. Whereas for the Mo compounds it is of the order of $\sim$150 meV, for the heavier W compounds increases up to $\sim$ 400 meV.  SOC also lead to a splitting of the conduction band at both, the band edge at the K point,\cite{KGF13} as well as at the secondary minimum Q which lies between the $\Gamma$ and K points of the BZ (indicated in Fig. \ref{Fig:1L}-\ref{Fig:Bulk} by a dashed vertical line). Notice that, since Q is not a high symmetry point of the BZ, the minima of the conduction band for bilayer and bulk materials do not lie exactly at the same point than for single layers. This is why the minima of the conduction band in Figs. \ref{Fig:2L} and \ref{Fig:Bulk} are slightly shifted with respect to the single-layer Q point. 

For monolayer samples and around the K and K' points, it is possible to assign a spin projection to the different Bloch states along the normal to the $MX_2$ plane. It is possible to define a sign for the SOC induced splitting of the band $n$ from the difference $\Delta^{\rm SO}_n(\bk)=\varepsilon_{n\uparrow}(\bk)-\varepsilon_{n\downarrow}(\bk)$. Time reversal symmetry implies that $\Delta^{\rm SO}_n(\bk)=-\Delta^{\rm SO}_n(-\bk)$, leading to splitting of opposite sign at the K and K' points. DFT calculations show, for the K point of the conduction band, a negative sign for $\Delta^{\rm SO}$ for the Mo compounds ($\sim -3$ meV for MoS$_2$ and $\sim -21$ meV for MoSe$_2$) and a positive sign for the W compounds ($\sim 27$ meV for WS$_2$ and $\sim 38$ meV for WSe$_2$).\cite{KGF13} Notice that the SOC splitting of the conduction band is larger for the Se compounds as compared to the S compounds. This is expected due to the heavier mass of  selenium as compared to sulfur. Since the orbital weight of the $p_x$ and $p_y$ orbitals of chalcogen atoms at the K point of the conduction band is of the order of $\sim 20\%$,\cite{CG13} first order processes associated to this $X$ atoms lead to a contribution to the SOC splitting, being more noticeable for the Se compounds. Recent calculations suggest that Coulomb interaction can modify the SOC split bands.\cite{FC14} 

\begin{table*}
\begin{tabular}{lcccccc}
\hline
\hline
& LDA      & GGA                     &FPLAPW    & HSE    & GW & Exp \\
\cline{2-7}
& \multicolumn{6}{c}{Monolayer} \\
\hline
\hline
MoS$_2$  
& 0.1901\cite{KH12} 
& 0.147\cite{SP13}/0.146\cite{R12}/0.147\cite{KGF13} 
& 0.148\cite{ZCS11} 
& 0.193\cite{R12}/0.188\cite{ZG13}      
& 0.164\cite{R12}/0.146\cite{CL12}/0.112\cite{MW13} 
& 0.150\cite{MH10}/0.170\cite{R12}  \\
WS$_2$ 
& - 
& 0.435 \cite{SP13}/0.425\cite{R12}/0.433\cite{KGF13} 
& 0.426\cite{ZCS11} 
& 0.521\cite{R12}   
& 0.456\cite{R12}   
&  - \\
MoSe$_2$ 
& -
& 0.183\cite{R12}/0.186\cite{KGF13}
& 0.183\cite{ZCS11}
& 0.261\cite{R12} 
& 0.26\cite{R12}
& 0.180\cite{ZS14}   \\
WSe$_2$ 
& -
& 0.461\cite{R12}/0.463\cite{KGF13}/0.43\cite{ZC13}
& 0.456\cite{ZCS11}
& 0.586\cite{R12}&0.501\cite{R12} &-\\
\cline{2-7}
& \multicolumn{6}{c}{Bilayer} \\
\hline
\hline
MoS$_2$  
& 0.1738\cite{MW13} 
& 0.17\cite{RNT11}
& - 
& 0.201\cite{ZG13}    
& 0.174\cite{CL12}/0.160\cite{MW13}
& 0.170\cite{MH10} \\
WS$_2$ 
& - 
& - 
& - 
& - 
& - 
& - \\
MoSe$_2$  
& - 
& - 
& - 
& - 
& - 
& - \\
WSe$_2$  
& -
& 0.43\cite{ZC13} 
& - 
& - 
& - 
& - \\
\cline{2-7}
& \multicolumn{6}{c}{Bulk} \\
\hline
\hline
MoS$_2$  
& 0.2201\cite{MW13}/0.258\cite{BP01}  
& 0.23\cite{RNT11} 
& 0.184\cite{ZG13}
& 0.238\cite{CL12}/0.2306\cite{MW13}  
& -
& 0.180\cite{MH10}/0.161\cite{BP01}  \\
WS$_2$ 
& -
& -  
& - 
& - 
& -
& - \\
MoSe$_2$
& 0.294\cite{BP01} 
& - 
& - 
& - 
& 0.175\cite{BP01} 
& -\\
WSe$_2$
& -
& -  
& - 
& - 
& - 
& -\\
\hline
\hline
\end{tabular}
\caption{Splitting at the valence band at K obtained in other works using different methods or functionals as LDA, GGA, all-electron full-potential linearised augmented-plane wave (FPLAPW), HSE and the GW approximation. All the energies are expressed in eV.}
\label{Tab:OthersDFTSplit}
\end{table*}

\section{Effect of strain}

The phonon structure of TMDs considered here is highly sensitive to strain. The Raman spectra of these compounds contain two main peaks which correspond to the A$_{1g}$ out-of-plane  mode, where the top and bottom $X$ atoms are moving out of plane in opposite directions while $M$ is fixed, and the $E_{2g}^1$ in-plane mode where the $M$ and $X$ atoms are moving in-plane in opposite directions.\cite{MW11,LR10} These phonon modes are red shifted with increasing temperatures, which might point the importance of anharmonic contributions to the interatomic potentials.\cite{LR13}  The $E_{2g}^1$ mode is very sensitive to applied strain. Applying uniaxial strain lifts the degeneracy of this mode, leading to red shifting and splitting into two distinct peaks for strain of just 1\%.\cite{CS13,CB13,HL13} The particular electronic structure and phonon modes of TMDs suggest that band structure engineering methods can be used for electronic and optoelectronic applications. This technique can be especially useful for TMDs which have been shown to sustain elastically deformations up to 11\% without breaking the material.\cite{CR12} 

\begin{figure}
\includegraphics[scale=0.15,clip=]{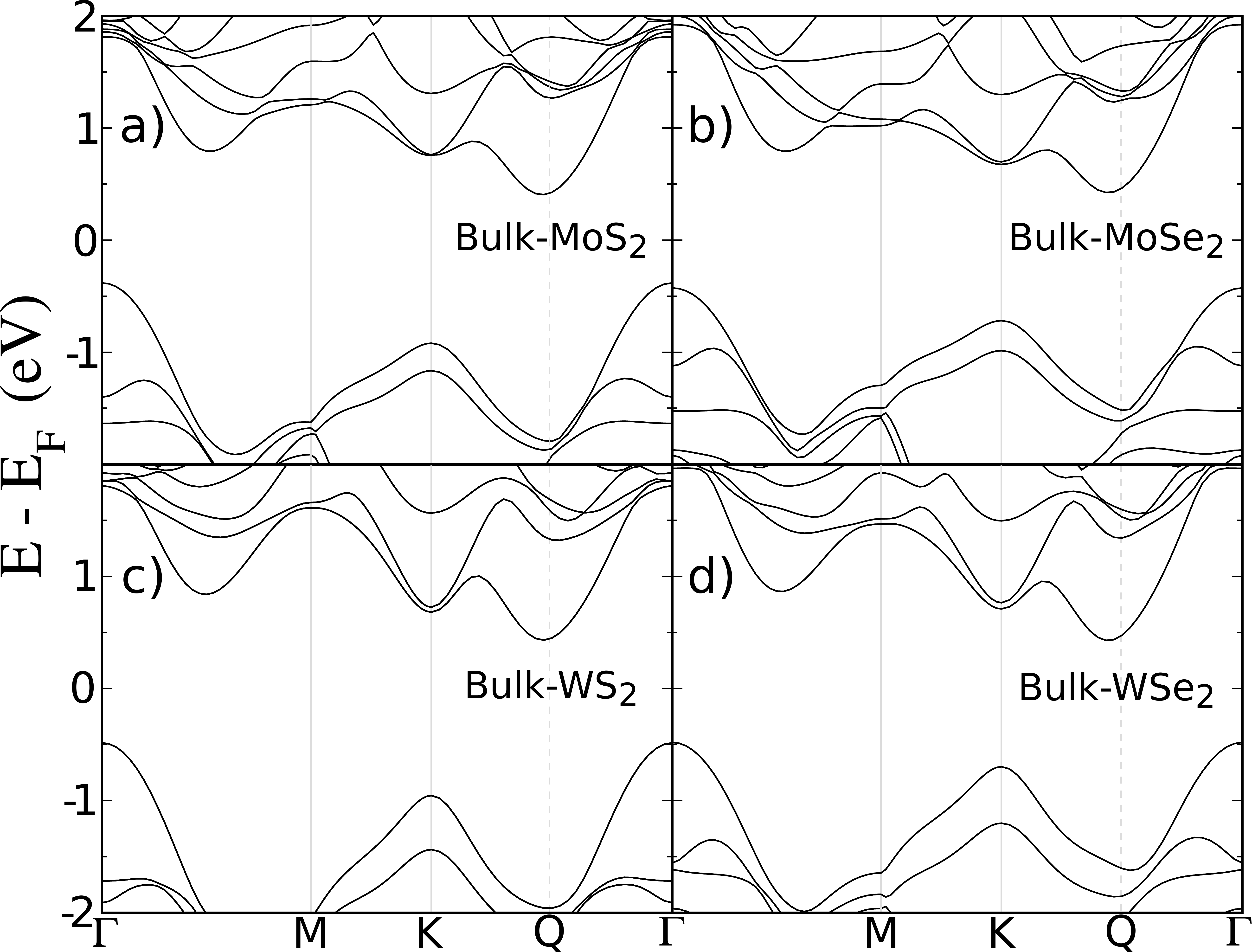}
\caption{Band structure of bulk MoS$_2$, MoSe$_2$, WS$_2$ and WSe$_2$
obtained from DFT calculations.
}
\label{Fig:Bulk}
\end{figure}

\begin{figure}
\includegraphics[scale=0.15,clip=]{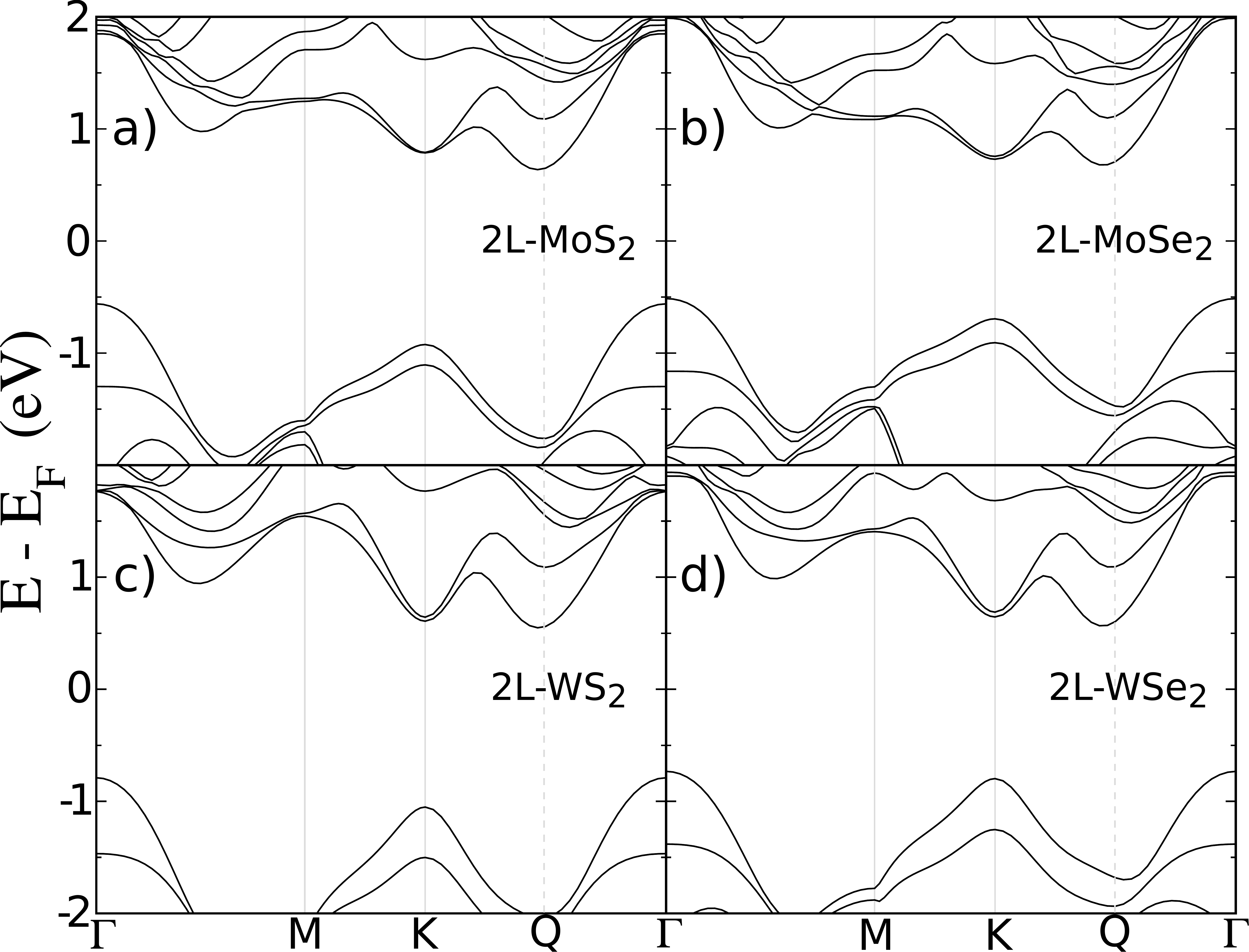}
\caption{Band structure of bilayer MoS$_2$, MoSe$_2$, WS$_2$ and WSe$_2$
obtained from DFT calculations.
}
\label{Fig:2L}
\end{figure}

Furthermore,
the SOC induces a finite band splitting in single-layer
systems also at the six inequivalent valleys at the Q point of the BZ,\cite{ZCS11,RG14}
with the corresponding entanglement of  spin/valley/orbital degrees of freedom.\cite{YC14}
At the microscopic level, we remind that
the main orbital character of the conduction bands at the Q point
is due to a roughly equal distribution of the $d_{x^2-y^2}$ and $d_{xy}$
orbitals of the transition metal $M$, and of the
$p_x$ and $p_y$ orbitals of the chalcogen atom $X$.
The rather large contribution from both $p$- and
$d$-orbitals leads to a  strong
hybridization between $X$ and $M$ atoms at this Q point of the BZ. This makes these states highly sensitive
to uniform and local strains and lattice distortions.\cite{CS13}
In fact, since the minimum of the conduction
band at Q becomes the effective band edge in bilayer and multilayer samples
(see Figs. \ref{Fig:2L} and \ref{Fig:Bulk}), 
this suggests that the states at  the minima of the conduction band
at the Q are good candidates for tuning
the spin/orbital/valley entanglement in these materials
by means of strain engineering\cite{CS13} or (in multilayer systems)
by means of electric fields.\cite{WX13} 

The application of forces across specific axes of the crystal structure induces strain in the sample which, at the same time, can be used to modify the band structure of TMDs. A reduction of the band gap can be achieved under uniaxial compressive strain across the $c$-axis of the crystal structure of $MX_2$.\cite{PV12} On the other hand, biaxial tensile strain makes the minimum of the conduction band at the Q point to move upward, while the conduction band at the K point moves downward with no significant effect in the valence band at the $\Gamma$ point. Furthermore, tensile strain is expected to lower the electron effective mass\cite{SY13} and consequently improve electron mobility, and could also lead to band degeneracy. Some of these theoretical results have been confirmed by strain experiments in single layer and multi-layer $MX_2$, which shows a change in the band gap up to $\sim 300$ meV per 1\% applied strain.\cite{HL13,CB13,CS13} Strain can be induced by depositing MoS$_2$ on hexagonal boron nitride (h-BN), which can lead to a direct-to-indirect gap transition, as predicted by first-principle calculations.\cite{HZ14}   Recent proposals suggest that time-reversal invariant topological phases in TMDs can be engineered by application of strain.\cite{COG14}
  
Of special interest is the use of strain engineering on $MX_2$ to create a broad-band optical funnel. This idea, proposed theoretically by Feng {\it et al.}\cite{FL12} and confirmed experimentally by Castellanos-Gomez {\it et al.}\cite{CS13} consists on continuously change the strain across a sheet of monolayer MoS$_2$, leading to a continuous variation of the optical band gap, which allows not only the capture of photons across a wide range of the solar spectrum, but also guidance of the resulting generated excitons toward contacts. 

\section{Effect of disorder}

Disorder in the lattice structure, such as adatoms or vacancies, can strongly modify the optical and transport properties of materials. In particular the mobility of TMDs is highly dependent on the screening environment and is limited by the presence of impurities in the samples.\cite{KTJ12,JH13,SH13,KTJ13} Experimental results show that the optical and transport properties of these materials can be influenced by the existence of defects in their chemical and structural composition.  Vacancies in $MX_2$ crystals trap free charge carriers and localize excitons, leading to new peaks in the photoluminescence spectra.\cite{TW13} Such vacancies can be created by means of thermal annealing and $\alpha$-particle \cite{TW13} or electron beam irradiation.\cite{ZI13} The optical spectrum of bilayer MoS$_2$ presents a broad peak at $\sim 1.77$ eV which  has been associated to impurities,\cite{WX13} whereas the mobility of multilayer samples has been shown to highly depend on the substrate and dielectric effects.\cite{BF13}   Line defects, which separate patches or islands where the layer direction is opposite to its surrounding, can lead to changes in the carrier mobility.\cite{ES13} Furthermore, short-range scatterers have been proposed as the main limitation for the mobility of chemical vapor deposition (CVD) grown single-layer MoS$_2$.\cite{ZA14,SE14} Recent scanning tunneling microscopy (STM) and scanning tunneling spectroscopy (STS) in combination with transport measurements suggest that the intrinsic $n$-doping in bulk MoS$_2$ is due to point defects which are consistent with S vacancies. Moreover, the significantly higher $n$-doping observed in thin films deposited on SiO$_2$ is extrinsic and has been attributed to trapped donors at the interface with the SiO$_2$ substrate.\cite{LA14} The controlled creation of defects could be exploited as a route to manipulate the electronic properties of these materials. 

On the theoretical ground, the effect of disorder has been first studied using DFT methods. \cite{AC11,MH11,WP12,KK12,GH13,ZZ13,ES13,LR13} The study of adsorption of adatoms and creation of vacancy defects in MoS$_2$ nanoribbons have shown  that a net magnetic moment can be achieved through the adsorption of Co adatoms to the non-magnetic armchair nanoribbons. Furthermore, apart from the spin-polarization, significant charges can be transferred to (or from) the adatom.\cite{AC11} Spin-polarized DFT calculations for MoS$_2$ antidots show that the net spin and the stability of spin states can be engineered by controlling the type and distance of internal nanoholes. The case of only S-terminated antidots is found that can exhibit a large net spin above room temperature.\cite{ZZ13}

The electronic and optical properties of single-layers of MoS$_2$ and WS$_2$ in the presence of vacancies have been also studied within a {\it real space} tight-binding (TB) model for large systems, containing millions of atoms, by solving the time-dependent Schr\"odinger equation, and by means of numerically exact Kubo formula calculations.\cite{YG14} Vacancies induce states in the middle of the gap whose energy depends on the specific vacant atoms, and the optical transitions involving the impurity bands lead to a background contribution in the  optical conductivity at low energies, which has been proposed to be behind the features observed in photoconductivity  experiments.\cite{MH10} $MX_2$ samples also show a significant asymmetry between electrons and holes,\cite{BF13} such that the DC conductivities and mobilities are larger for holes. When comparing the $M$S$_2$ compounds, it is found higher mobilities for $p$-doped WS$_2$ than for MoS$_2$.\cite{YG14}

\section{Spin/valley/layer coupling}\label{Sec:Coupling}

In single layers of $MX_2$, the lack of inversion symmetry as well as the SOC leads to a break in
spin degeneracy along both, the valence and conduction bands, and also causes valley Hall effect where carriers flow into opposite transverse edges upon application of an in-plane electric field.\cite{CF12,ZC12,LS13,MM14} In addition, time reversal symmetry of the single layers of $MX_2$ together with non degeneracy of the spin energy bands lead to a coupling of spin and valley states.\cite{XY12} This results in valley-dependent optical polarization selection for individual valleys. Since time reversal symmetry forces opposite spin-splitting at each valley, such effect can be used to control the valley polarization by means of optical helicity,\cite{MH12} such that the K valley would correspond to optical selection rules of a certain helicity as well as carriers of a fixed spin, while the K' valley would correspond to opposite conditions. This makes possible to control carrier spin as well as carrier confinement within a specific valley with circularly polarized light, an effect that has been observed experimentally.\cite{MH12,ZC12,CF12} Since inter-valley scattering is in principle forbidden due to the breaking of spin degeneracy, this suggests long spin lifetimes in single layers of $MX_2$. As a matter of fact, spin lifetimes larger than 1 ns have been determined experimentally,\cite{MH12} in agreement with theoretical calculations for intravalley SOC mediated spin relaxation in MoS$_2$.\cite{OR13} In the absence of defects, spin relaxation of the carriers is also possible due to flexural deformations of the samples.\cite{OGF13}

New interesting features are present in bilayers of $MX_2$, which consists on a stack of two single layers in-plane rotated by $180^\circ$ 
with respect to each other, bound by means of weak van der Waals
interactions.
The inter-layer hopping of electrons between different layers leads
to a strong modification of the band structure, driving a transition
from a direct gap semiconductor in single-layer systems
to an indirect gap semiconductor in bilayer compounds, as it can be seen by comparing Fig. \ref{Fig:1L} and \ref{Fig:2L}.
The inter-layer hopping, which
links mainly the $p$ orbitals of the chalcogen atoms $X$ of different
layers,\cite{CG13} lead to a splitting of the maximum of the
valence band at the $\Gamma$ point, which becomes the effective
valence band edge, as well as a splitting of the minimum of the
conduction band at the Q point, which becomes the absolute minimum
of the conduction band.\cite{RG14}

Contrary to single-layer {\em MX}$_2$, bilayer {\em MX}$_2$ presents
point-center inversion symmetry.\cite{WX13,GY13,ZC13}
Therefore, the band structure of bilayer $MX_2$
remains spin degenerate even in the presence of SOC.
However, since the SOC does not couple orbitals of
different layers, each single band preserves a finite entanglement between
spin, valley and the layer index.
Such spin-valley-layer coupling can be observed at the K point of the valence band,\cite{GY13}, as well as at the conduction band.\cite{RG14}  This last case can be of special interest for slightly electron-doped bilayer $MX_2$, whose Fermi surface
presents six pockets centered at the inequivalent Q valleys of the BZ,
and no pockets at the K and K' valleys. Of special interest are the families of TMDs with stronger spin-orbit interaction, like WS$_2$ and WSe$_2$, for which the SOC can be larger than the inter-layer  hopping, enhancing the spin/layer/valley entanglement.
Then, although inversion symmetry forces each Fermi pocket to be spin
degenerate,
the layer polarization makes that each layer
contributes with opposite spin in alternating valleys.
This effect can be useful for {\it valleytronics} devices. 
The importance of spin and valley states was proven by optical probes in bilayer MoS$_2$, obtaining a reduction of the photoluminescence by more than 20 times, and hole spin lifetime was 3 orders smaller than those observed in single layers of MoS$_2$.\cite{MH12}  The control and tuning of circularly polarized photoluminescence from bilayer MoS$_2$ can be achieved with the application of a gate voltage, which breaks inversion symmetry due to the electric field, as it has been shown by recent experiments.\cite{WX13}

\section{Excitons}

The additional {\it spin-like} quantum numbers discussed in Sec. \ref{Sec:Coupling} play an important role in the physics of excitons in TMDs. The reduced dielectric screening in monolayer and few layer samples suggest that strong excitonic effects should appear in these materials. In fact, the existence of highly stable neutral and charged excitons has been proven experimentally.\cite{Ross_2013,MS13,Jones_2013}  In those experiments, optically excited electrons and holes are bound together by means of Coulomb interaction. Contrary to excitons in conventional semiconductors as GaAs, for which the excitons form at the $\Gamma$ point of the BZ, excitons in TMDs occur at the K and K' points of the BZ, leading to the so called valley excitons, which open new opportunities to manipulate and control the valley index by means of optical probes, as we have discussed in the previous section.  The large SOC splitting of the valence band at the K point leads to two excitonic features in the photo-absorption spectrum,\cite{MH10} usually denoted as the A and B excitons. On the other hand, the binding energy of the charged excitons (usually called {\it trions}) is unexpectedly large (18 meV for MoS$_2$,\cite{MS13} and 30 meV for both MoSe$_2$,\cite{Ross_2013} and WSe$_2$\cite{Jones_2013}), pointing out that Coulomb interactions are very strong in these families of layered TMDs. 

A large exciton binding energy ($\sim$0.32 eV) and a significant deviation from the conventional hydrogenic model, typically used to describe the Wannier excitons in inorganic semiconductors, has been found in the full sequence of excited (Rydberg) exciton states in monolayer and few-layer WS$_2$, measured by optical reflexion experiments.\cite{CH14} A theoretical microscopic model that considers the modification of the functional form of the Coulomb interaction due to the nonlocal nature of the effective dielectric screening has been successfully applied to explain these unusually strong electron-hole interactions.\cite{CH14}

To tackle this problem using first principle methods, one needs to go beyond DFT calculations since it is well know that Kohn-Sham energies do not correspond to quasiparticle energies. A better route is to consider $GW$ approximation in conjunction with Bethe-Salpeter equation (BSE) to consider the two-particle excitations.\cite{CL12,R12} Other possibility is to use an effective mass model (parametrized by {\it ab initio} calculations), and including appropriate screening of the interactions for quasi-2D semiconductors.\cite{BHR13} The separation between the excitons is directly related to the strength of the SOC splitting of the valence band, and excitation energies in the range 1 to 2 eV have been predicted,\cite{R12} suggesting a potential application of TMDs for optoelectronic devices in the near-IR to the red regime. In Table \ref{Tab:ExcitonsBE} we list the binding energies of the neutral excitons obtained theoretically from different approximations, as well some experimental values  that have been reported in the literature.

\begin{table*}
\centering
\begin{tabular}{lcccccccc}
\hline
\hline
         & \multicolumn{2}{c}{Monolayer} & & Bilayer & & \multicolumn{2}{c}{Bulk} \\
 & Effective model & GW & & GW & & Exp. & GW \\
\hline
MoS$_2$   
&0.54\cite{BHR13}&0.897\cite{CL12}/1.03\cite{R12}/1.1\cite{KK12}/0.5\cite{FL12}/0.54\cite{SY13} & 
& 0.424\cite{CL12} 
& 
& 0.08\cite{KK12}
& 0.13\cite{KK12}/0.025\cite{CL12}  \\
WS$_2$    
& 0.47\cite{BHR13}
& 0.91\cite{R12}
& 
& - 
& 
& -
& -\\
MoSe$_2$  
& 0.50\cite{BHR13} 
& 1.04\cite{R12}/0.54\citep{SY13}           
&  
&  -
&     
& 0.07\cite{KK12}    
&  0.11\cite{KK12} \\
WSe$_2$   & 0.42\cite{BHR13} 
&  0.90\cite{R12}            
&        
&  -
& 
& -
& - \\
\hline
\hline
\end{tabular}
\caption{Binding energies (in eV) for the neutral excitons obtained from different approaches and from experiments, for monolayer, bilayer and bulk TMDs.}
\label{Tab:ExcitonsBE}
\end{table*}

\section{Superconductivity}

Another interesting aspect of the TMDs is the appearance of a superconducting transition with a critical temperature that strongly depends on the carrier density. A superconducting dome, similar to that observed for the layered cuprate, has been experimentally observed in the temperature-carrier density phase diagram of MoS$_2$,\cite{YI12,TT12} in which the samples were doped by a combination of liquid and solid gating. For the optimal doping  $n\sim 1.2\times 10^{14}{\rm cm}^{-2}$ the critical temperature is of $T_c\sim 10.8$~K. 

Theoretically, the origin of superconductivity in heavily doped MoS$_2$ was first studied in Ref. \onlinecite{RCG13}, by considering the role of both electron-electron and electron-phonon interactions. The estimations for the strength of the different contributions to the effective coupling suggest that superconductivity in MoS$_2$, under the experimental conditions of Refs. \onlinecite{TT12} and \onlinecite{YI12}, is likely to be induced by the electron-electron interaction. The significant short-range repulsion between carriers at the conduction band allows for a superconducting phase induced by the electron-electron interaction with a nontrivial structure, where the gap acquires opposite signs in the two inequivalent pockets of the conduction band. On the other hand, DFT calculations for the phase diagram of TMDs suggest that phonon mediated superconductivity is also possible for some range of electron doping, and a charge density wave is also proposed to exist for even higher carrier concentrations.\cite{RHW14} Recently it has been suggested that spin-triplet $p$-wave superconductivity can be stabilized by Rashba SOC in MoS$_2$.\cite{YML14} Finally, the superconducting proximity effect and the Andreev reflection in $MX_2$ superconducting-normal (S/N) hybrid junction with n- (p-)doped S and p-doped N regions has been investigated in Ref. \onlinecite{MRA14}, finding that the strong SOC enhances the Andreev conductance of the MoS$_2$-based S/N structure relative to its value in the corresponding structure with gapped graphene.

\section{Conclusions}

In conclusion, we have discussed the electronic properties of semiconducting TMDs $MX_2$, where $M={\rm Mo, W}$ and $X={\rm S, Se}$. DFT band structure calculations, including the effect of SOC, have been used to discuss the differences between the single layer, bilayer and bulk compounds. The splitting of the bands due to SOC has been discussed in the whole BZ, analyzing the role of the transition metal and the chalcogen atoms at each relevant point of the band structure. We have further reviewed the effect of strain in the samples, discussing how strain engineering is a good route to manipulate and tune the electronic and optical properties of those compounds. Disorder and imperfections in the lattice structure lead to the creation of impurity states in the middle of the gap, which contribute to the photoconductivity. Finally, we have discussed the superconducting phase which has been also experimentally observed in the TMDs, and which present a superconducting dome in the temperature-carrier density phase diagram, which resembles to that of the cuprate superconductors. The different mechanisms proposed to explain the superconducting pairing have been reviewed.  

\acknowledgments

R.R., M.P.L.-S. and F.G. acknowledge financial support from MINECO, Spain,
through grant FIS2011-23713, and the European Union,
through grant 290846. R. R. acknowledges financial support
from the Juan de la Cierva Program (MINECO, Spain).
E.C. acknowledges support from the
European project FP7-PEOPLE-2013-CIG ``LSIE\_2D''
and Italian National Miur Prin project 20105ZZTSE. 
J.A.S.-G. and P.O. ackowledge support from Spanish MINECO
(Grants No. FIS2012-37549-C05-02 with joint financing by FEDER Funds from the European Union, and
No. CSD2007-00050) and to Generalitat de Catalunya (2014 SGR 301). J.A.S.-G. was supported by an FPI Fellowship from MINECO. The authors thankfully acknowledges the computer resources, technical expertise and assistance provided by the Red Espa\~nola de Supercomputaci\'on.

\bibliography{BibliogrGrafeno}

\end{document}